\documentclass[11pt]{amsart}

\usepackage[margin=1in]{geometry}

\usepackage{amsmath}
\usepackage{graphicx,nicefrac}
\usepackage[colorinlistoftodos]{todonotes}
\usepackage[colorlinks=true, allcolors=blue]{hyperref}
\usepackage{boldline,multirow,hyperref}
\usepackage{color,colortbl}

\newtheorem{theorem}{Theorem}

\newtheorem*{claim}{Claim}
\newtheorem*{remark}{Remark}

\title[Representational Baseline]{Locating the representational baseline:\\
Republicans in Massachusetts\\}
\author[VRDI]{Moon Duchin, Taissa Gladkova, Eugene Henninger-Voss, 
Ben Klingensmith,\\ Heather Newman, and Hannah Wheelen}
\date{\today}
\begin{document}

\begin{abstract}
Republican candidates often receive between 30\% and 40\% of the 
two-way vote share  in statewide elections in Massachusetts.
For the last three Census cycles, MA has held  
9-10 seats in the House of Representatives, 
which means that a district
can be won with as little as 6\% of the statewide vote.
Putting these two facts together, one may be surprised to learn 
that a Massachusetts Republican has not won a seat in the U.S. House
of Representatives since 1994.  We argue that 
the underperformance of Republicans in Massachusetts is not attributable to 
gerrymandering, nor to the failure of Republicans to field
House candidates, but is a structural mathematical feature
of the distribution of votes.   For several of the elections studied here,
there are more ways of building a valid districting plan than there are particles
in the galaxy, and every one of them will produce a 9--0 Democratic
delegation.
\end{abstract}

\maketitle

\vspace{-.25in}
\section{Introduction}

{\em Gerrymandering} is the practice of using the formation of 
districts to create a representational advantage for some 
subsets of the population, or to favor certain kinds of 
candidates.  In recent years, gerrymandering has received increasing
levels of attention and public indignation.  
There are essentially two indicators that are taken by the 
public and by many commentators as red flags for gerrymandering:  {\em bizarre shapes}
and {\em disproportional outcomes}.
For instance, the enacted 113th Congressional districting plan
in Pennsylvania contained a notorious district nicknamed
``Goofy kicking Donald Duck,'' whose contorted shape was taken by many 
as prima facie evidence of redistricting abuse.  
Under this map, Pennsylvania elections exhibited
nearly 50-50  splits in party preference, while Republicans 
 held 13 out of 18 seats, or over 72\% of the House representation.
While there is indeed compelling evidence that Pennsylvania
was gerrymandered in a partisan manner \cite{md-report}, this 
fact is not established by either shapes or disproportions alone.
In this paper, we show that there can also exist benign and
structural obstructions to securing representation that 
have to do with not just the number of votes but how they 
are distributed around the state.

This paper is framed to study a riddle about 
Republican voting patterns in Massachusetts:  
{\em why is 1/3 of the vote proving insufficient to secure even
1/9 of the representation?}  
We use a mix of empirical analysis with real voting
data and 
experiments with generated voting data to answer
the riddle.
We show that {\em uniformity 
itself can block desired representational outcomes for a group in the numerical
minority} (like Republicans in Massachusetts), considering
both the numbers and the geometry.
Though this is mathematically obvious when taken 
to an extreme,  exhibiting actual voting 
patterns with this level of uniformity is a novel
finding.

Massachusetts is one striking case in point, but the broader message is that once the rules have been set,
it becomes a scientific question to study the breadth of outcomes left available to the districters.  This case describes
a surprising limitation on the power to control the representational outcome.  In other cases, there will 
be surprisingly wide latitude, or simply a baseline in a non-intuitive range.  We argue that it is only legitimate
to compare an observed partisan outcome against the backdrop of actual possibility.

\subsubsection*{Numerical uniformity}
We use the phrase ``numerical uniformity'' to describe a situation in which the vote shares across the building-block units
are extremely consistent.  
In Section~\ref{sec:numerical} we  examine the numerical
distribution of votes in 13 statewide
elections in Massachusetts, showing that for five of them, 
the numbers alone make it literally impossible to build a R-favoring
collection of towns or precincts with enough population to be a Congressional
district.  Because this type of  analysis is run on the numbers only, 
this result is very strong:  no district-sized grouping can be formed,
even without requiring contiguity, compactness, or any other
spatial constraint on districting.  The reason is that elections with
in which Republicans are locked out exhibit extremely {\em low variance}
in the town-  and precinct-level voting results.
In particular, even in some elections in which a Republican received 30-40\% of the 
overall vote, his vote share (note: they have all been men) rarely exceeded 50\% in any precinct,
leaving not enough R-favoring precincts 
to assemble into a grouping of the size of a congressional district.

\subsubsection*{Geometric uniformity}
On the other hand, ``geometric uniformity'' would describe a situation in which one's partisan preference
does not correlate strongly to position within the state, reflected in the absence of partisan enclaves or clusters.
In  Section~\ref{sec:geometric} we will add a spatial component to our analysis. 
Even when it is numerically feasible to collect enough precincts to form an R-favoring
district, the precincts may not be spatially located in 
such a way that this can be accomplished in 
 a connected (i.e., contiguous) fashion.  
We first show visualizations that illustrate the lack of a 
Republican enclave in the low-variance elections.  
These graphics suggest that there is low correlation between
location and partisanship in these Massachusetts elections.
To corroborate this, we 
compute clustering scores (which measure segregation
of Republican votes from Democratic votes).
We find that the actual vote distributions have clustering
levels that are similar to those that would be observed
if sprinkling the Republican votes by drawing randomly from a uniform distribution around the state.
This supports the conclusion that geometric uniformity is making a secondary contribution 
to the partisan underperformance.

\bigskip

In short, the conclusion is that extreme representational outcomes
are not always attributable to gerrymandering,
nor to overly clustered arrangements of voters from either party, 
which have sometimes been claimed to force a heavily clustered party to win districts with wastefully high majorities (via ``packing''). 
On the contrary, in this case Republicans are locked out of
representation because they are {\em insufficiently} clustered:
here, the main
factor responsible for the lockout of Republicans is actually
that the minority party is distributed {\em too uniformly} 
around the state, both numerically and geometrically.
Generally, counterintuitive limitations on representation can emerge
from a complicated interplay of the numerical and 
spatial distribution of voter preferences.  The effects
on representation
of the distribution (and not just the share) of votes 
is a difficult mathematical question and is richly worthy of 
further study.

While public observers may expect proportional representation as a matter of fairness,
even seasoned political scientists often measure fairness in terms of other representational indices.
For instance, the efficiency gap, or $EG$, is sometimes described as measuring parity of wasted votes, but is fundamentally measuring
whether the seat share $S$ is close to $2V-1/2$, where $V$ is the vote share.
The efficiency gap, $EG=2V-S-1/2$,  is thought to signify a possible gerrymander when its magnitude is more than $8\%$.  
But the Massachusetts data contain five actual vote distributions (Pres 00, Pres 04, Sen 06, Pres 08, Sen 08)
for which even an omniscient redistricter with the honorable goal of $EG=0$ could not succeed:
 not a single one of the quintillions of possible 9-district plans has an efficiency gap below $11\%$ in any of those five races. 
This shows that finding a reasonable baseline to decide when gerrymandering has occurred is a subtler problem than has so far been appreciated 
in the public discourse or the political science literature.

\newpage
\subsection{Data}\label{sec: data}

Massachusetts is made up of 351 jurisdictions known as towns (also written in 
some places as townships or municipalities), which has not changed over the timespan covered here.  
In this language, cities are a subset of towns.  
Towns do not overlap, and they completely cover the state.  Each town is 
subdivided into some number of precincts, ranging in number from 2166 in 
2002 to 2174 in 2016 according to the Secretary of State database \cite{secstate}.  Small changes to precincts are common between elections.
In 2016, 125 towns were not
subdivided (the town equals 1 precinct), and at the other 
extreme, Boston was made up of 255 precincts, followed by Springfield
with 64.  
Note that precincts are similar but not identical to Voting Tabulation Districts or VTDs, which are proposed by the Census Bureau every 
ten years as recommended precincts \cite{vtd tigerline} and are adopted in a slightly modified form in Massachusetts.  
The Secretary of State's office provided us with a VTD shapefile that reflects the state's intended precincts in 2010.
The Census provides a town shapefile.

After the 2010 Census, the number of Congressional delegates apportioned to MA dropped from 10 to 9 because the state's population growth did not keep 
pace with the country's.

In the tables below, the cast vote data comes from the Secretary of State's website \cite{secstate}.  
They offer town-level election
results back to the year 2000 and earlier, but precinct-level
results only back to 2002.  For population numbers, town-level population was retrieved from the Census API directly. 
Census 2000 population figures were used for elections taking place 2000--2010, and
Census 2010 for 2010--2016. 
The Secretary of State's shapefile included VTD population numbers, but because it did not perfectly match with the precincts in the 
voting tabular data, 
population was aggregated up from census blocks to VTDs,  and these populations were verified using NHGIS data. 
We then prorated election data from towns for each election into these VTDs by assigning each VTD the proportion of each candidate's town-wide
vote that corresponds to that unit's proportion of the town's area.   We note that there is no name-matching used in this process.
To assign VTDs to districts in the currently enacted plan, we used the TIGER/Line shapefiles from the 113th Congress,
rounded onto towns and precincts by areal allocation.

All of our data, together with scripts needed to run the various algorithms
described here, can be found in the public github repositories of the 
Voting Rights Data Institute \cite{github chain,github mass}.

\subsection{Setup choices: Election data, number of districts, smallest units, constraints}

In order to illustrate this effects of uniformity observed
in real voting data, we
run this feasibility analysis on election results from 13 
Presidential and U.S. Senate elections in Massachusetts. 
We note that  Congressional election results are not considered here because many of the recent races are uncontested. 
For example, in the 2016 U.S. House election, 5 out of 9 districts had no Republican who filed to run \cite{ballotpedia}. 
Therefore, two-way vote share analysis would not be meaningful for these races.\footnote{U.S. Senate 
voting patterns are well known  
to be more closely correlated with Congressional preferences than Presidential votes, but that is somewhat beside the point for this analysis, 
which is focused on the range of representational outcomes that are possible for given observed partisan voting patterns.}  
We will also choose to analyze the seat share possible out of nine Congressional districts for the sake of consistency, even though
our timespan of electoral data includes a period over which the apportionment varied between 9 and 10.
Neither decision blunts the impact of the findings, which study the extent to which patterns in real voting data
can restrict the range of representation that is possible for a group in the numerical minority. 

In the numerical feasibility section we will only require that districts hew close to the standard of equal population and that they 
are made of whole units, sometimes towns and sometimes precincts.  
Contiguity of districts and other shape constraints will only be discussed in the geometric section of the paper.
Because of the importance of real voting data for this analysis, we must use precincts as the smallest 
building blocks, since that is the smallest level at which vote returns are available.
In practice, the 2011 Congressional plan held 2119 precincts intact while splitting 32, which means that fewer than $1.5\%$ were split.    

Using towns or precincts as unsplittable building blocks does have some precedent in law and practice.  
As a historical matter, the state Constitution of Massachusetts did 
require in Article XVI that state councillors be elected from contiguous districts that keep intact towns and city wards \cite{const-MA}, 
but this system of councillors is now obsolete.
There is a still-active contiguity requirement for state legislative districts, 
and a rule to preserve towns as much as is ``reasonable,''
but no formal contiguity or unit-preservation requirement for congressional districts. 
In fact, only 23 states have a contiguity requirement for congressional districts, while 49 require  contiguity for legislative districts.  
Nonetheless congressional district contiguity is essentially universal in practice.\footnote{District contiguity can be made somewhat complicated by 
water and by smaller geographic units that are themselves disconnected, but these issues are relatively easy to resolve in Massachusetts.
Districting rules may be found in the state constitution \cite{const-MA} and at \url{http://redistricting.lls.edu/states-MA.php}.}

\subsection*{Acknowledgments}
We gratefully acknowledge the Bose Research Grant and PI Justin Solomon
for major support of the Voting Rights Data Institute.
We thank Gabriela Obando and William Palmer at the MA 
Secretary of State's office for their help collecting and interpreting 
data. Thanks also to Jowei Chen for sharing a dataset that approximates
precinct-level vote counts in 2000, to Gary King for extremely useful feedback,
and to Max Hully and Ruth Buck for excellent data support.

\section{Arithmetic of Republican underperformance}\label{sec:numerical}

\begin{figure}[ht]
\includegraphics[width=.49\textwidth]{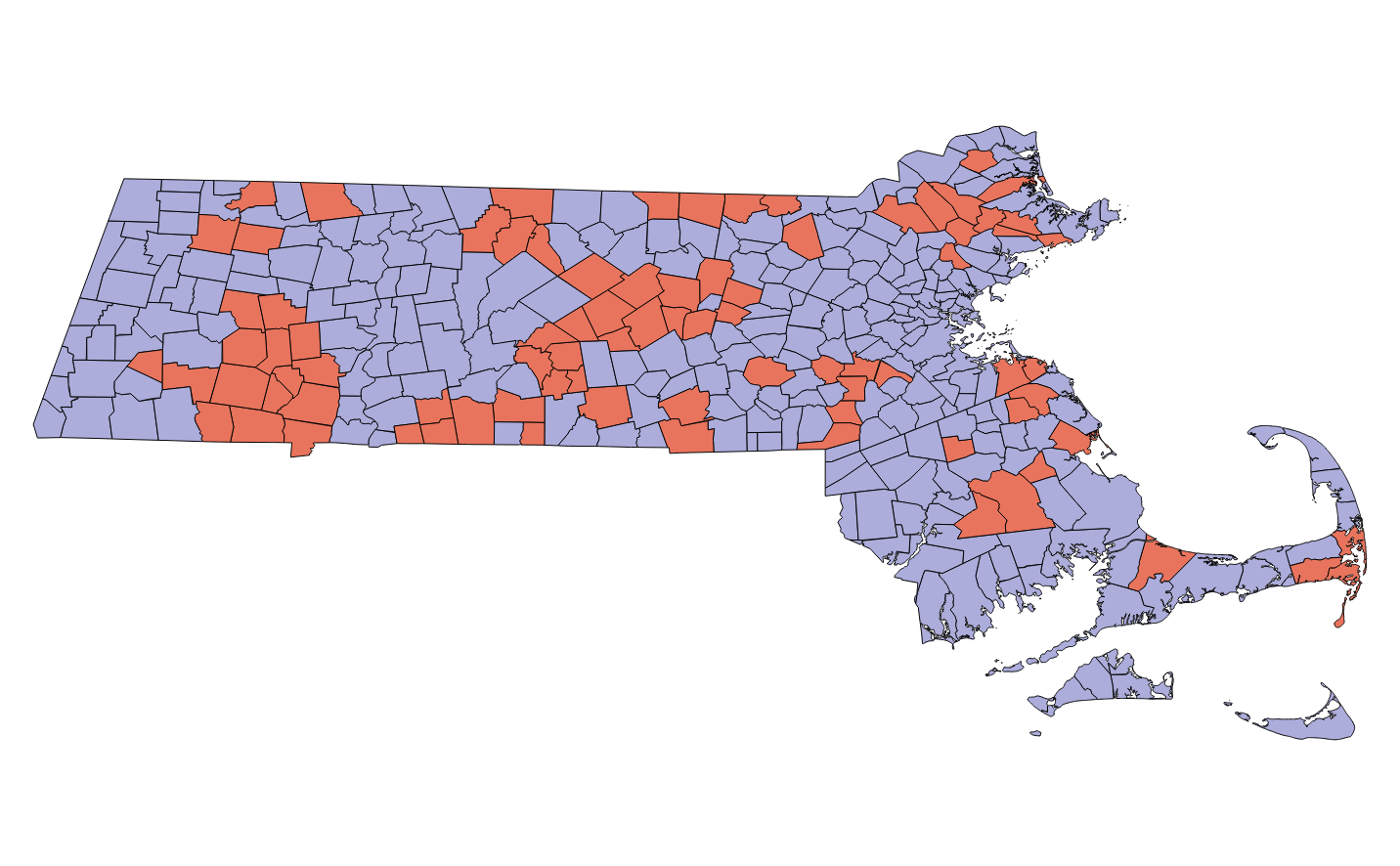}
\includegraphics[width=.49\textwidth]{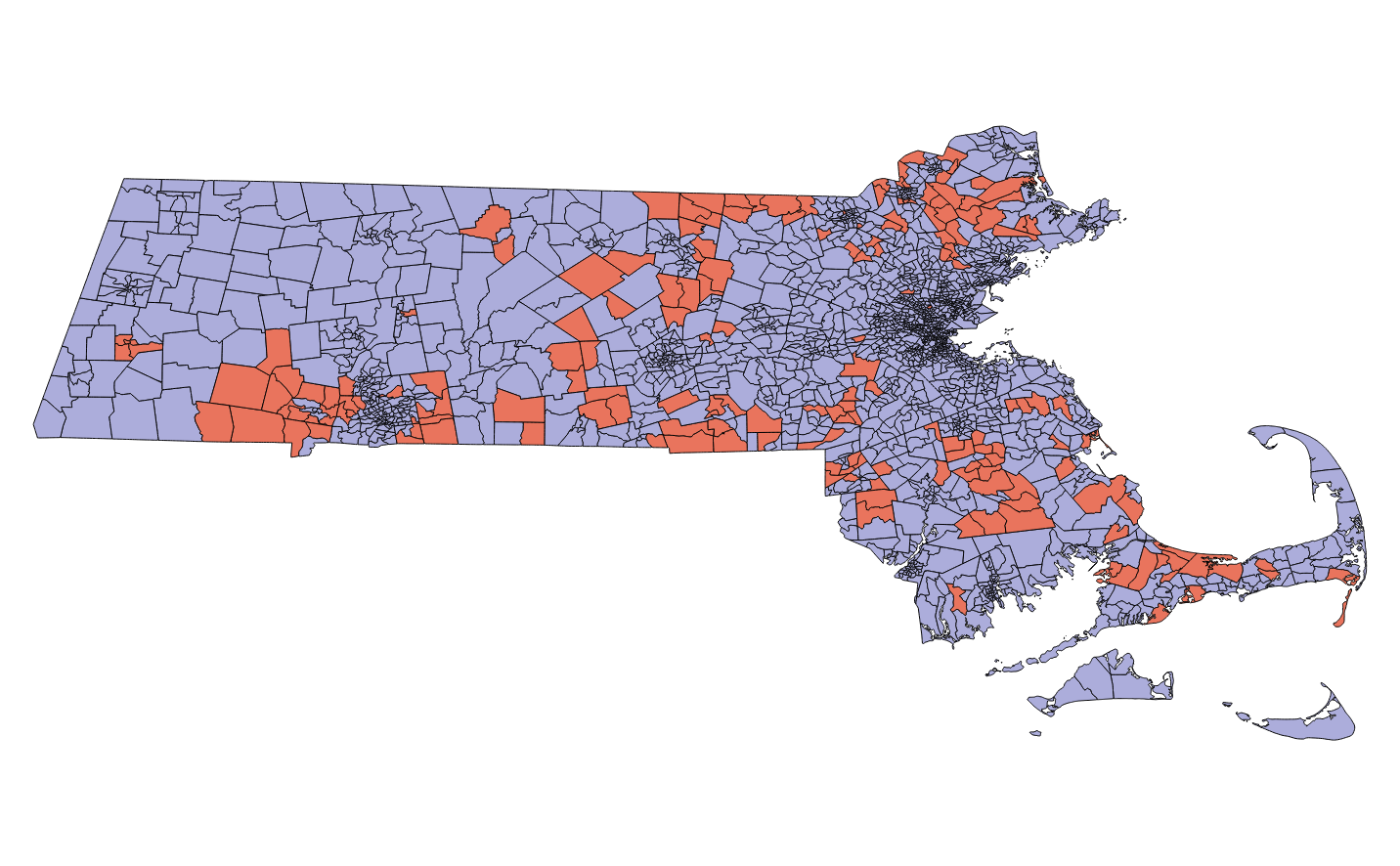}
        \caption{This figure shows the district-sized collection of towns most favorable to George W. Bush in the 2000 Presidential race (left), and the collection of precincts most favorable to Kenneth Chase
        in the 2006 Senate race (right).  These ``districts'' still preferred Gore and Kennedy, respectively.\label{gwbush}}
    \end{figure}

In this section, we 
describe a method to determine 
 theoretical bounds on the number 
 of districts with a Republican majority, given only
 the geographical units, their population, and their vote totals
 for D and R candidates in a particular election.
For this part of the analysis we impose no spatial constraints at all;
we do not even require contiguity, but would allow
a district constructed out of an arbitrary collection of 
towns or precincts from around the state.  
We show, for example, that {\em even though George W. Bush
received over $35\%$ of the two-way vote share against
Al Gore, it is mathematically impossible
to construct a collection of towns, however scattered, with at least 10\% of the population and where Bush received more collective votes than Gore}.   (See Figure~\ref{gwbush}.)


\begin{table}[htbp]
\begin{tabular}{|l||c|c|c|c|c|} \hline\hline
Election &  R Share& 
\multicolumn{2}{|c|}{R Share by Town} &\multicolumn{2}{|c|}{R Share by Precinct}\\ \hline 
\multicolumn{2}{|c|}{}&mean&variance&mean&variance\\
\hline
\rowcolor{pink}
Pres 2000 &  35.2\%   & 39.70\% &.0074 &-- & -- \\
\rowcolor{pink}
Sen 2000& 25.4\%$^*$  & 29.15\% &.0044 &--  & -- \\
\rowcolor{pink}
Sen 2002 & 18.7\%     & 20.29\% &.0020&17.43\%  & .0028\\
\rowcolor{pink}
Pres 2004 &  37.3\%   & 40.00\%   &.0093 &34.53\% & .0140 \\
\rowcolor{pink}
Sen 2006 &  30.6 \%   & 33.24\% &.0077 &27.59\%  & .0119 \\
\rowcolor{pink}
Pres 2008  & 36.8\%   & 39.00\% &.0117 &33.80\%  & .0181\\
\rowcolor{pink}
Sen 2008& 32.0\%      & 34.40\% &.0094 &28.87\% & .0142\\
 Sen 2010 & 52.4\%     & 53.79\% &.0202&47.71\%  & .0310\\
 Pres 2012 & 38.2\%    & 41.06\% &.0146&34.91\%  & .0228 \\
 Sen 2012& 46.2\%      & 49.20\%  &.0169&42.70\% & .0275\\
 Sen 2013& 44.9\%      & 48.89\% &.0217&41.89\%  & .0312\\
 Sen 2014& 38.0\%      & 41.15\% &.0141&34.28\%  & .0206\\
 Pres 2016& 35.3\%     & 40.18\% &.0165&33.12\%  & .0236\\
 \hline\hline
\end{tabular}
\caption{Statistics of Republican two-way vote share in 13 statewide
elections in Massachusetts.  Lower-variance elections are
marked in red.
(* Libertarian vote share included with R in 2000 Senate race) }
\label{tab: variance}
\end{table}

\begin{remark}[The Boston Effect]
Note that in Table~\ref{tab: variance} the town-level 
mean R share reliably overshoots the statewide R share,
while the precinct-level mean errs in the other direction. Recall that there are 351 towns in the 2016 election, 
subdivided into 2174 precincts. Boston is composed of 255 precincts; Springfield has 64; and most other towns have fewer than 25 precincts, with 125 towns (more than a third) having only one.   
This means Boston is an outlier in size, and it
is also an outlier in the lopsidedness of its Democratic 
voting majority. (In the 2016 Presidential election, Boston had only a 14.7\% R two-way vote share.) 

The town-level averaging underweights Boston because
it is weighted equally to tiny towns like Gosnold (population 75). 
The precinct-level 
results overweight Boston because its average precinct population is under 2500,
lower than the statewide average of over 3000.  (Exact figures vary year to year.)
This accounts for the direction of error in 
the mean of each statistic relative to the 
statewide (naturally population-weighted) share.
\end{remark}

As the table illustrates, the elections from 2000 to 2008 had consistently lower variance
in their town- and precinct-level vote shares than can be observed since 2010.  
Below, we will connect that to the representability of Republicans across these elections.

\subsection{Numerical feasibility of R districts} \label{sec: numerical feasibility}
Let's first review the limitations on the power of gerrymanderers that are produced by the numbers alone.
We begin with very simplified algebraic bounds.
In an abstract districting system with equal vote turnout in its districts,
if Party X  receives share $0\le V\le 1$ of the vote, its possible seat shares are constrained to a range,
with the actual outcome depending on how the votes are distributed across the districts.
At its most ruthlessly efficient, Party X could in principle have barely more than half of the vote in certain districts and no vote in the others,
thus earning seat share up to $2V$, or twice its vote share.  At minimum, a party with less than half of the vote 
can be shut out entirely by having less than half of each district; if Party X has more  than half of the vote,
then its opponent has a vote share of $1-V$ and a maximum seat share of $2(1-V)=2-2V$,
so the  minimum seat share for Party X is $1-(2-2V)=2V-1$.  
For example, a party with $40\%$ of the vote can get anywhere from $0-80\%$ of the seats,
while a party with $55\%$ of the vote can get anywhere from $10-100\%$ of the seats.
This naive analysis would project that districters could in principle arrange for Beatty voters in the 2008 Senate
race to convert their $32\%$ of the votes to $0-64\%$ of the seats.  

But the naive analysis does not take into account constraints introduced by the fixed number
of districts, by the variation in turnout, or by the discreteness of the building blocks.  
The feasibility analysis in this section does account for all of those factors.  
Table~\ref{tab: town feasible} shows that in Ed Markey's 2013 special election to the Senate,
his opponent's pattern in obtaining 
$38\%$ of the vote could not have earned him any more than three district wins out of nine,
no matter how the districts were drawn, despite the naive bounds that suggest up to six district wins
could have been possible.
And even more strikingly, though Jeff Beatty earned nearly
a third of the vote against Kerry in the Senate race of 2008, Beatty voters in that distribution are actually
locked out of representability entirely.  The actual observed turnout patterns, and the effect 
of the mandate to build districts out of intact precincts, have lowered Beatty's ceiling from 5 districts out of 9
all the way to zero.  Smaller building blocks should mean more flexibility, but 
shrinking the building blocks from towns to precincts didn't in this case help 
Beatty at all.

Here is our method for measuring feasibility in our setup.  
Suppose that the ideal district size (state population divided
by number of districts) is denoted by $I$.
Then we will declare that it is
{\em numerically feasible} for a party to get $n$ seats
in a certain election
if there exists a collection of units (towns or precincts)
with population at least $nI$ and in which that 
party has a majority of the two-way vote share.
A feasibility bound for the party is the largest
such $n$ that has been demonstrated.

By contrast, we will say that it is {\em numerically infeasible}
for a party to get $m$ seats in a given election if there 
is proven to be no collection with population at least $mI$ and a majority
for the party.  An infeasibility bound is the smallest such $m$
that has been demonstrated.

We use a simple sorting algorithm to get feasibility and 
infeasibility bounds for the elections considered here,
presenting the results in Table~\ref{tab: town feasible}.
Often, but not always, the algorithm will produce tight bounds,
in the sense that the infeasibility bound is one more than 
the feasibility bound.\footnote{It is possible that the feasibility bound actually overstates the 
number of districts that can be built with a majority for the designated party---because the collection
of size $nI$ may not be splittable into $n$ appropriate collections of size $I$---but any infeasibility bound 
reflects a mathematically proven impossibility, which drives all the conclusions in this section.}

Our procedure is simply
to greedily create the largest R-majority collection possible from 
the chosen geographic units
(in our case, towns or precincts) by including them in order
of Republican margin per capita:  
$$\delta/p =(\# \hbox{R votes} - \# \hbox{D votes})/(\hbox{census
population of unit}).$$

The proof supporting this test of feasibility is shown in the appendix, \S\ref{sec: greedy proof}.

We will carry out the analysis below fixing the number of districts at 9 throughout, which 
is the Congressional apportionment at the current time.  
This means that ideal district size is 
$I=705,455$ for races before 2010 and $I=727,514$ for races after 2010.  

\begin{table}[htbp]
\begin{tabular}{|l||c|c|c|c|c|c|c|} \hline\hline
Election & D Candidate--R Candidate&  R Share& Seat Quota &
\multicolumn{2}{|c|}{R Feas/Infeas} &\multicolumn{2}{|c|}{D Feas/Infeas}  \\ \hline 
\multicolumn{3}{|c|}{}&(9 seats)&town&prec&town&prec \\
\hline
\rowcolor{pink}
Pres 2000 & {\bf Gore}--Bush & 35.2\% & 3.2&0/1 & --- &9/- & --- \\
\rowcolor{pink}
Sen 2000& {\bf Kennedy}--Robinson/Howell & 25.4\%$^*$& 2.3 &0/1& --- &9/- & --- \\
\rowcolor{pink}
Sen 2002 & {\bf Kerry}--Cloud & 18.7\%& 1.7
 &0/1 & 0/1 &9/- & 9/-\\
\rowcolor{pink}
Pres 2004 & {\bf Kerry}--Bush  & 37.3\% & 3.4
 &1/2 & 1/2&9/- & 9/- \\
\rowcolor{pink}
Sen 2006 & {\bf Kennedy}--Chase & 30.6 \%& 2.8
 &0/1 & 0/1 &9/- & 9/-\\
\rowcolor{pink}
Pres 2008 & {\bf Obama}--McCain & 36.8\% & 3.3
 &1/2 & 1/2&9/- & 9/- \\
\rowcolor{pink}
Sen 2008& {\bf Kerry}--Beatty & 32.0\%& 2.9
 &0/1 & 0/1 &9/- & 9/-\\
 Sen 2010& Coakley--{\bf Brown} & 52.4\%& 4.7
 &9/- & 9/- & 8/9 & 8/9 \\
 Pres 2012& {\bf Obama}--Romney & 38.2\% & 3.4
 &3/4 & 3/4 & 9/- & 9/- \\
 Sen 2012& {\bf  Warren}--Brown & 46.2\% & 4.2
 &7/9 & 7/8 &  9/- & 9/- \\
 Sen 2013& {\bf Markey}--Gomez & 44.9\% & 4.0
 &7/9 & 7/8 &  9/- & 9/- \\
 Sen 2014& {\bf Markey}--Herr & 38.0\% & 3.4
 &3/4 & 3/4 & 9/- & 9/- \\
 Pres 2016& {\bf Clinton}--Trump & 35.3\% & 3.2
 &2/3 & 3/4 & 9/- & 9/- \\
 \hline\hline
\end{tabular}
\caption{If districts were to be made out of towns 
or out of precincts, with no regard to shape or even connectedness,
how many R or D districts could be formed?
Feasibility and infeasibility bounds are shown in this table.
Low-variance elections (see previous table) are marked in red.
Election winners shown in boldface; R share is with respect
to 2-way vote; seat quotas are proportional share of 9 seats.\label{tab: town feasible}}
\end{table}

We can make several observations from the table. Moving to finer 
granularity of building blocks did not have any impact on the feasibility bounds for most elections.  
In two cases (Sen 2012 and Sen 2013), the precinct level bounds are sharper.  In both 
cases, a Republican-performing grouping of towns can be made with size $7I$ but 
our method produces an inconclusive result about a grouping of size $8I$.  With
precincts, we find that the uncertainty is eliminated and a grouping of size $8I$ is impossible.
The 2016 Presidential election is the only one for which the finer granularity has shifted the 
feasibility bounds.  It is not possible to find scattered towns totaling three districts' worth of population
which collectively favor Trump over Clinton, but it becomes possible if precincts are the building blocks.
So in that case, it becomes narrowly possible to achieve proportional representation for Trump voters;
note, however, that this still falls far short of the {\em seven} Trump districts that the simple analysis would have predicted to be
accessible by extreme gerrymandering.

\subsection{Numerical uniformity: The role of variance}

In statistics, the {\em mean} of a set of numerical data
records its average value, and the {\em variance}
(or second central moment) tells you how spread
out the values are around this mean.  
We claim that variance in the vote share of a minority
group (here, Republicans)
can be a primary explanatory factor for 
poor representational outcomes in districting.
At one extreme, this is obvious:  if the variance
is zero, then 
the preferences in the state are completely uniform, and 
every single unit has the same 35\% (say) 
of Republican votes.  
In this case, we can easily see that districting has no 
impact at all:  every possible district will also have
35\% R, and so will be won by Democrats.

Notably, the Gore/Bush election in 2000 had a  two-way R vote share of 35.2\% and results in zero possible R-majority districts. Meanwhile the Clinton/Trump election had a nearly identical
35.3\% R vote share but produces the possibility for 
as many as {\em two} districts with a Trump majority.

    \begin{figure}[ht]
    	\includegraphics[width=0.32\textwidth]{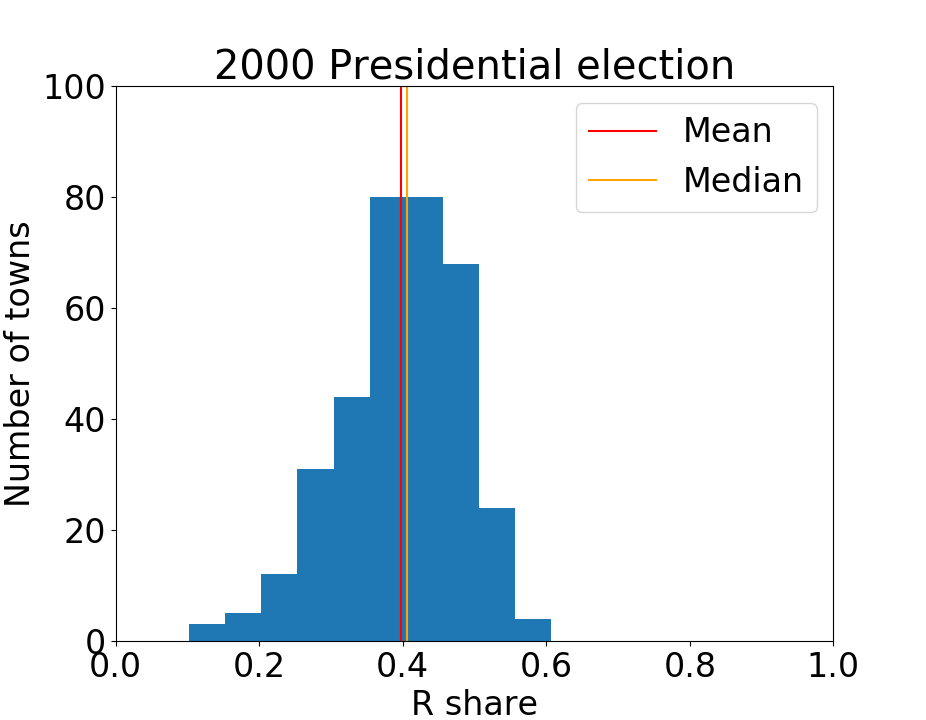}
        \includegraphics[width=0.32\textwidth]{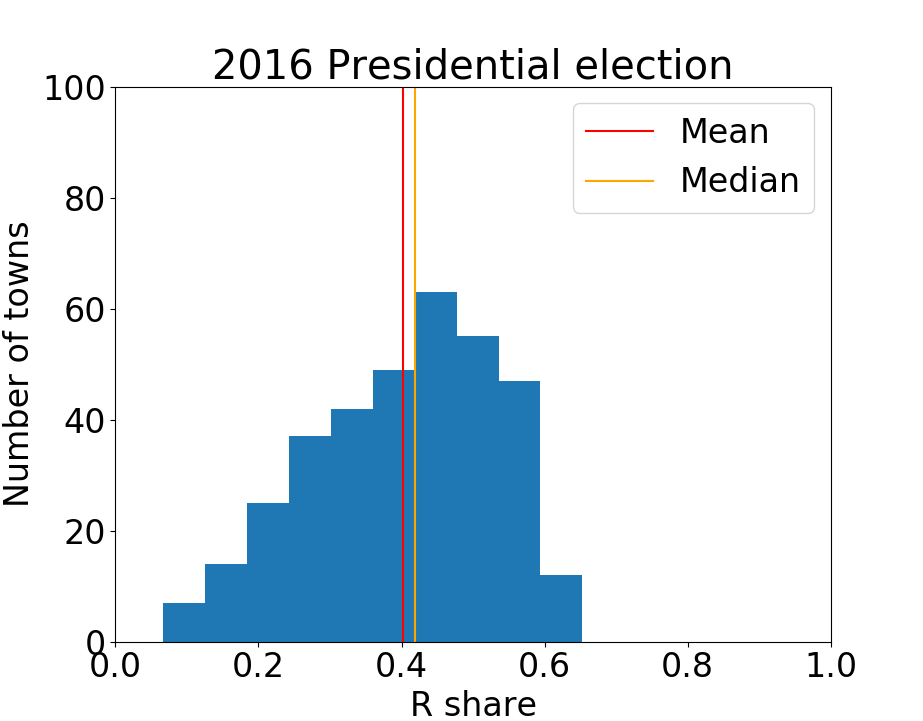}
\caption{These histograms show the distribution 
of Republican vote share by town in the 2000 and 
2016 MA Presidential contests, illustrating 
elections with very nearly the same mean but 
different levels of variance.  These two elections have town-level variance $.0074$ and $.0165$, respectively.\label{fig: real distributions}}
    \end{figure}

The fundamental impact of variance should be clear from the 
figures.  A low-variance election with a minority of R votes
 may have very few units with 
R share over $.5$, which are precisely the 
building blocks needed to form an 
R-majority district.  

Looking back to  Table~\ref{tab: town feasible} 
corroborates this finding:
7 out of 13 elections exhibit a mathematical
impossibility of representation or fall at least two seats short of proportionality---completely 
independent of the choices made by districters.  These
are precisely the seven elections in which the vote totals
show lower variance, both at the town level and the precinct 
level.   In five of the elections, this 
effect is so pronounced that the minority party is completely
locked out of any possibility of  representation.

\subsection{Varying variance}
\begin{figure}[ht]
    	\includegraphics[width=0.32\textwidth]{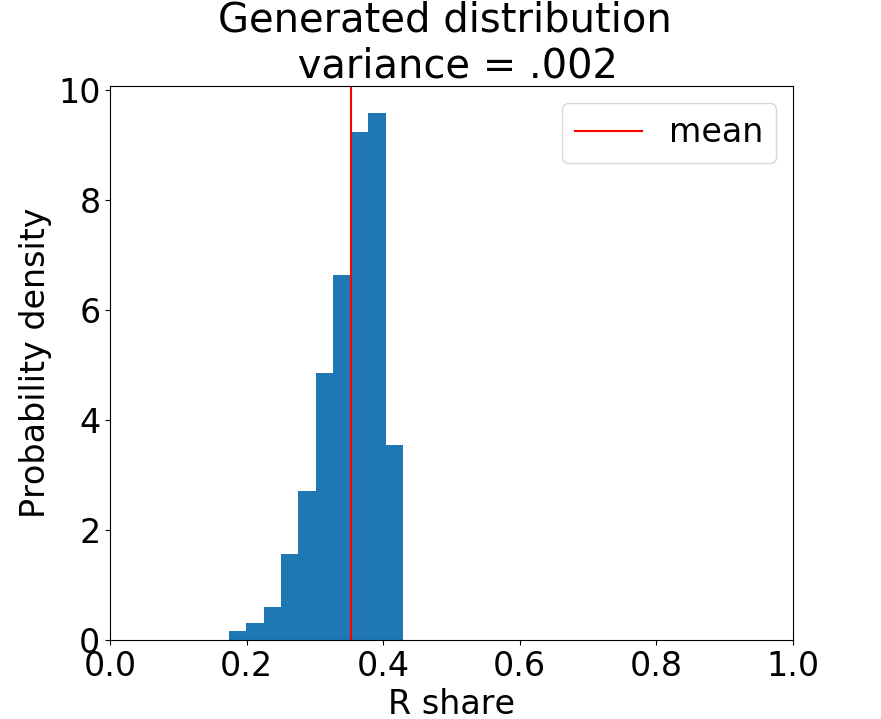}
        \includegraphics[width=0.32\textwidth]{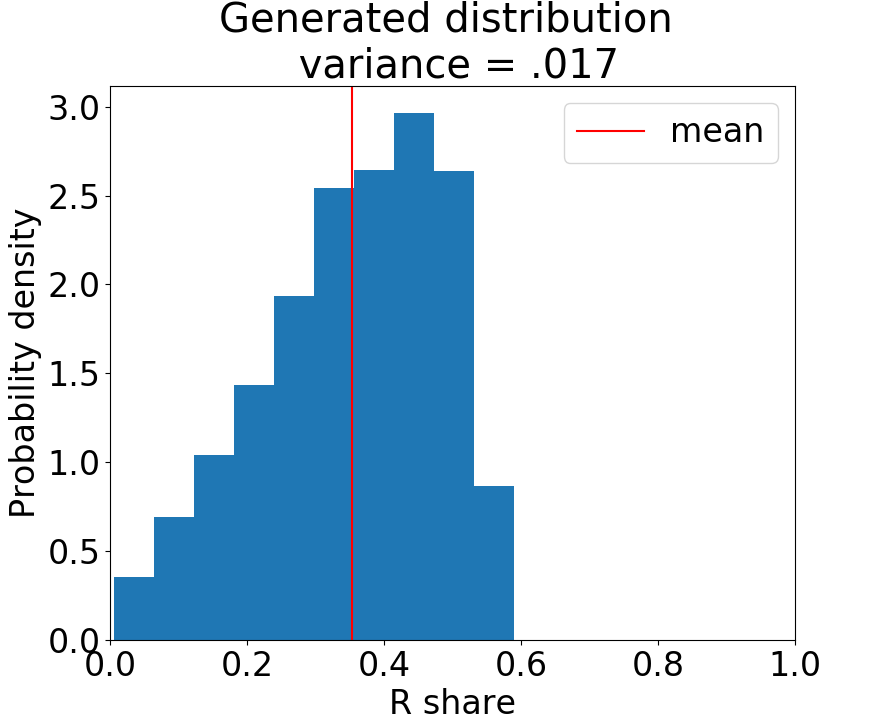}
        \includegraphics[width=0.32\textwidth]{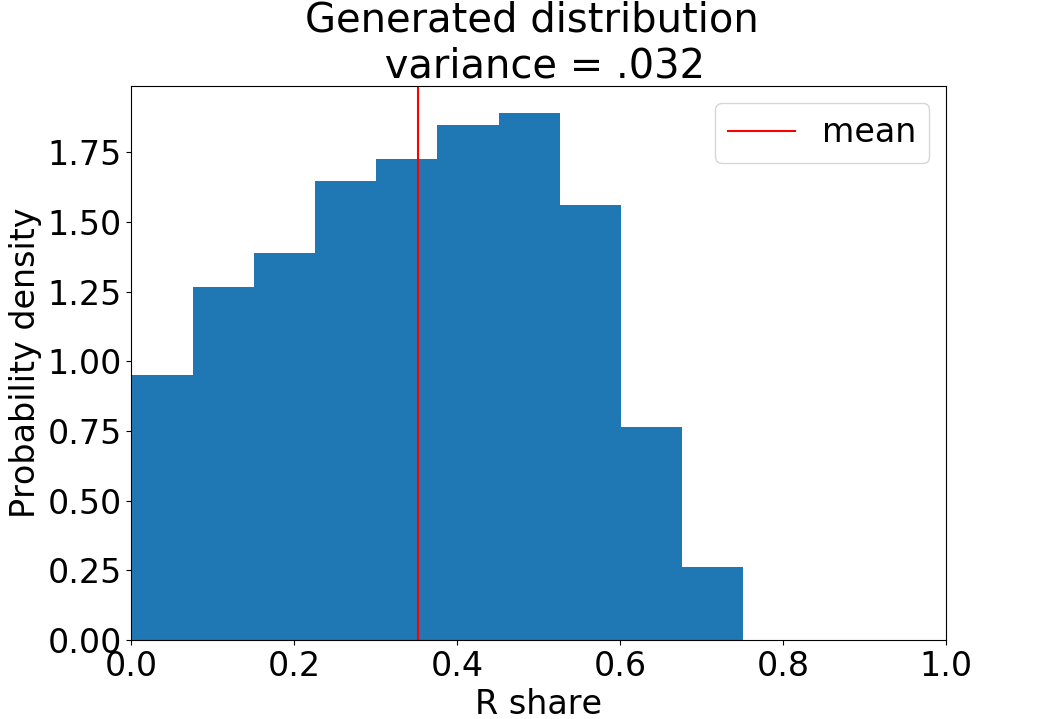}
        \caption{Skewed
 truncated normal distributions are shown here with the same
mean and different variance as the observed results.  These were used to generate election data to test the hypothesis 
that vote datasets with higher variance would achieve higher levels of numerically feasible representation.\label{fig: generated distributions}}
    \end{figure}

To account for these outcomes, 
we generated datasets with similar mean vote share to the 2000 and 2016 Presidential elections, 
adjusting the variance of R-share per unit while maintaining  
voter turnout and population at actual levels.
We assigned R two-way vote shares chosen from a truncated skewed normal distribution with a set mean of 
35.25\% (the average of the Gore/Bush and Clinton/Trump R vote share) and  variances ranging from 
0.0020 to 0.0320, 
covering the range actually observed in Table~\ref{tab: variance}.\footnote{We used the scipy python library 
{\sf skewnorm.rvs} function  to generate random numbers  from a skewed normal distribution with the chosen location, scale, and shape variable. 
{\em Truncation} means that any value outside of the $[0,1]$ range was replaced by another value drawn from the same distribution.  This truncation process changes the mean and variance of the distribution being produced, so we ran it iteratively, adjusting the mean and variance  until the desired parameters were produced.  Throughout, a shape variable of $-8$ was selected to best capture the observed distributions in historical elections. The resulting distributions can be seen in Figure~\ref{fig: generated distributions}.} 
From those datasets, we reran our procedure to produce bounds on the number of possible R seats.

\begin{figure}[ht]
    	\includegraphics[width=.41\textwidth]{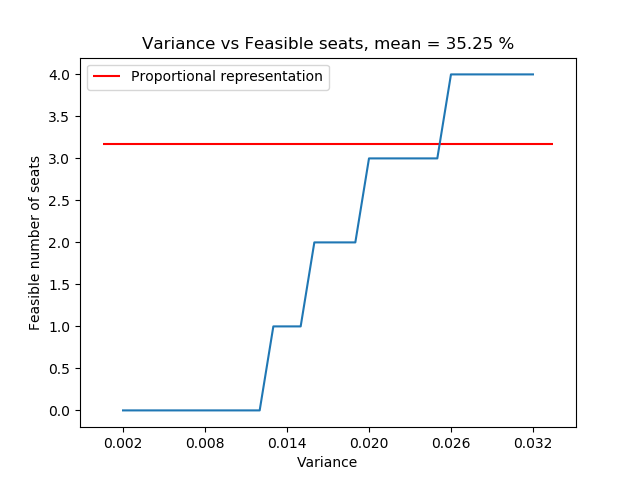}
        \caption{Higher-variance datasets reliably produce greater numbers of feasible seats,
        even with the vote share held constant.  This figure shows the results of three trials with the protocol
        described above; the results are visually indistinguishable.\label{fig: simulation results}}
    \end{figure}

The results, plotted in Figure~\ref{fig: simulation results}, strongly corroborate the hypothesis that 
feasible representation is directly controlled by variance
in vote share.  In fact, a high enough variance can be seen to make it numerically feasible to overperform 
proportionality.

\section{Geometry of Republican underperformance}\label{sec:geometric}
We now consider the spatial aspects of the vote distribution with respect to the possibilities
for district formation.

\subsection{Lack of Republican enclaves}

Compounding the numerical effects described above is the spatial scatter of the areas preferring 
Republicans in Massachusetts.  To illustrate this, consider forming a grouping of  towns by collecting them in order of their R margin
per capita $\delta/p$, as above, until the collection is large enough to be a valid district.  The result is a dramatically discontiguous
assemblage spanning nearly the full state.  A similar pattern can be observed in 2006 Senate returns.

In fact, very few of the building blocks shown in the picture are R-favoring at all.  Only 31 out of 351 towns had a G.W. Bush majority in 2000, and the largest
Bush-favoring collection of towns only has population 426,304, well short of ideal district size of over 700,000.  (Its Bush majority has a one-vote margin.)  
Similarly, only an astonishing 9 of 2166 precincts in 2006 record a Chase majority.

\subsection{Clustering}
The voting data used here makes it possible to test whether, in addition to increased variance,
the election results after 2010 exhibit more spatial clustering than before.  
To assess this we use an index called a {\sf capy} (or clustering propensity) score, which resembles 
assortativity scores in network science.  (See \cite{capy} for a comparative survey of scores of 
clustering and segregation.)

The geographical units that make up a jurisdiction have populations
of different sizes and compositions.  
In geographical unit $v_i$, we use  $x_i$ and $y_i$ to denote the populations from group X and group Y
in that  unit.
We  record the X population data as an integer-valued vector
$\mathbf x = (x_1,\dots,x_n)$ recording each unit's population, and likewise write $\mathbf y$ for the Y
population figures.  If unit $v_i$ has a shared boundary of positive length with unit $v_j$, we write $i\sim j$.
Then let 
$\displaystyle \langle \mathbf x,\mathbf y \rangle :=\sum_{i} x_iy_i + \sum_{i \sim j} x_iy_j + x_jy_i $.  
The idea is that $\langle \mathbf x,\mathbf y \rangle$ is a close approximation to the number of 
individuals of X type living next to an individual of Y type, either in the same geographical
unit or in neighboring units.\footnote{This approximation approaches equality as the populations get large.
For details, see \cite{capy}.}  
With this, we define
$$H(\mathbf x,\mathbf y):= \frac 12 \left(   
\frac{ \langle \mathbf x,\mathbf x \rangle} { \langle \mathbf x,\mathbf x \rangle +  \langle \mathbf x,\mathbf y \rangle}  
+
\frac{ \langle \mathbf y,\mathbf y \rangle} { \langle \mathbf y,\mathbf y \rangle +  \langle \mathbf x,\mathbf y \rangle} 
\right).$$
By construction, this score varies from $0$ to $1$ and 
measures the tendency of each of the two kinds of population to live next to another member of 
their own group, rather than the other.  
In a sufficiently large network, a perfectly uniform distribution where the $x_i$ and the $y_i$ were constant
would earn a score approaching $H=1/2$, and a perfectly clustered distribution where the $x_i=0$ in one region and the $y_i=0$ in the complementary
region would tend towards $H=1$.

\begin{table}[htbp]
\begin{tabular}{|l||c|c|c|c|} \hline\hline
Election & R  Share &  uniform $H$   & observed $H$ & clustered $H$ \\
\hline
Pres 2000 &  35.2\%    & .5001 & .5135 & .9456 \\
Sen 2000& 25.4\%$^*$    & .5000 & .5063& .9374 \\
Sen 2002 & 18.7\%      & .5001 & .5035& .8982\\
Pres 2004 &  37.3\%       & .5000 & .5182& .9351\\
Sen 2006 &  30.6\%         & .5001 & .5171& .9537\\
Pres 2008  & 36.8\%         & .5000 & .5210& .9591\\
Sen 2008& 32.0\%        & .5000 & .5181& .9513\\
 Sen 2010 & 52.4\%        & .5001 & .5329&  .9587\\
 Pres 2012 & 38.2\%      & .5000 & .5243& .9268\\
 Sen 2012& 46.2\%      & .5000 & .5272& .9597\\
 Sen 2013& 44.9\%       & .5002 & .5366& .9492\\
 Sen 2014& 38.0\%       & .5001 & .5276& .9557\\
 Pres 2016& 35.3\%        & .5000 & .5344& .9480\\
 \hline\hline
\end{tabular}
\caption{Clustering scores for Republican versus Democratic voters at the town level in each of the 
elections discussed in this paper.  We show the score $H=H(R,D)$ for a uniform trial, the observed votes, and a highly clustered distribution
of voters, each with the statewide share that corresponds accurately to the given election. 
The numbers are truncated (not rounded) after four decimal places.\label{tab: capy}}
\end{table}

Table~\ref{tab: capy} shows the observed $H(R,D)$ clustering results for Republican compared to Democratic voters.
For each election, we create two comparison points by experiment:  the {\em uniform $H$} score is the highest score recorded in 30 trials in which Republican voters 
were scattered randomly under a uniform distribution until reaching the statewide R share observed in that election.
The {\em clustered $H$} score is produced by applying a dynamical step that moves votes into a configuration with higher tendency 
for neighbors to have the same vote.\footnote{This is called the Ising model, and code can be found in our github repo \cite{github mass}.}
As a general matter, we  see that the $H$ scores closely resemble the uniform trials, and that there is no significant trend in the $H$ scores over time.  
In some cases, there are interesting comparisons, such as in comparing the Presidential outcomes
in 2000 and 2016---there, we can see that Trump voters are appreciably more clustered than Bush 
voters were.  We conclude that clustering may have a secondary effect on representability, 
but in a direction that runs counter to the conventional wisdom:  the prospects of the minority 
party for representation get better, not worse, when the voters are more tightly spatially clustered.

We note also that there is a one-way relationship between numerical and geometric uniformity:  
if there is low variance in observed partisan shares by unit, then all units tend to have the same shares, so there is necessarily no spatial pattern to 
partisan preference.  However, high variance in partisan share can occur in a way that is strongly spatially patterned (such as if there are 
pronounced enclaves) or in a way that is not (such as if there is a checkerboard pattern of strong support for each party).  
The findings here strongly support a conclusion that numerically uniform vote patterns create obstructions to representation for a group
in the numerical minority.  Further work is needed to study the spatial determinants of representability in the high-variance case.

\bigskip

In closing, we reiterate the main lesson of this simple study:
the range of possible representational outcomes  under valid redistricting
is controlled  by the  numerical and geometric/spatial distribution of voter preferences, and by the local rules of redistricting, in an extremely complex way
that one-size-fits-all normative ideals fail to capture.  The mathematical challenges of identifying the representational baseline are considerable, 
but there is  significant recent progress in that direction.
Any meaningful finding of gerrymandering must be demonstrated against the backdrop of valid alternatives---under the 
constraints of law, physical geography, and political geography that are actually present in that jurisdiction. 

\newpage
\bibliographystyle{alpha}

\begin{thebibliography}{99}
\bibitem{github chain} Metric Geometry and Gerrymandering Group, {\em Markov Chain Monte Carlo python package},\\
\url{https://github.com/mggg/GerryChain}.  Developed at Voting Rights Data Institute 2018.
\bibitem{github mass} Metric Geometry and Gerrymandering Group, {\em Massachusetts election data repository},\\
\url{https://github.com/gerrymandr/Massachusetts_underperformance}.  
\bibitem{capy} E. Alvarez, M. Duchin, E. Meike, and M. Mueller, 
{\em Demographic segregation and electoral representation}, preprint.
\bibitem{md-report}M. Duchin, {\em Outlier analysis for 
Pennsylvania}, February 2018.  LWV vs. Commonwealth of Pennsylvania Docket No. 159 MM 2017.
\bibitem{const-MA}Massachusetts Constitution.  \url{https://malegislature.gov/laws/constitution}
\bibitem{ballotpedia}Ballotpedia, U.S. House of Representatives elections in Massachusetts, 2016. \\  \url{https://ballotpedia.org/United_States_House_of_Representatives_elections_in_Massachusetts,_2016}
\bibitem{secstate}Massachusetts Secretary of State, {\em Massachusetts Election Statistics}.  \url{http://electionstats.state.ma.us}
\bibitem{vtd tigerline}U.S. Census Bureau, TIGER/Line Shapefile, 2012, 2010 state, Massachusetts, 2010 Census Voting District State-based (VTD) \\
\url{https://catalog.data.gov/dataset/tiger-line-shapefile-2012-2010-state-massachusetts-2010-census-}
\url{-voting-district-state-based-vtd}
\end{thebibliography}

\vspace{1in}
\section{Appendix: Rigorous feasibility bounds \label{sec: greedy proof}}
Suppose you have a list of units with corresponding populations 
$p_i$ and R margins $\delta_i$ (number of R votes minus
number of D votes).  Re-index so that they are ordered from greatest to least by margin per capita: 
$$\nicefrac{\delta_1}{p_1} \ge \nicefrac{\delta_2}{p_2}\ge \dots \ge
\nicefrac{\delta_n}{p_n}.$$
We will call a collection of units $S$ a {\em grouping}, and let
$p(S)$ and $\delta(S)$ be its population and R margin, found
by summing the $p_i$ and $\delta_i$ for its units.
Let $D_k$ be the grouping indexed by $\{1,\dots,k\}$.
Let $K$ be the smallest integer $k$ for which $\delta(D_k)\le 0$. 
This means that  $D_{K-1}$ has a collective
R majority,
but if you add the $K$th unit you get a grouping $D_K$ 
that fails to have an R majority.

\begin{theorem} With the notation above, let $M$ be any
positive integer.
\end{theorem}
\subsection*{Case 1} $M \leq p(D_{K-1})$.
There {\bf exists} an R-majority grouping of size at least $M$. 

\subsection*{Case 2} $p(D_{K-1}) < M \le p(D_{K})$.
Inconclusive: such a grouping may or may not exist.

\subsection*{Case 3} $p(D_{K}) < M$.
There {\bf does not exist} an R-majority grouping 
of size at least $M$. 

\begin{proof}
In Case 1, it is clear that a Republican grouping can be created, because $D_{K-1}$ is a Republican-majority grouping of sufficient size.

We present examples to illustrate that Case 2 is inconclusive.
\begin{center}
\begin{tabular}{cV{2.5}c|c|c|c}
$i$ & $r_i$ & $d_i$ & $p_i$ & $\nicefrac{\delta_i}{p_i}$ \\ \hlineB{2.5}
 1 &   8   &   0   &   8    &  1\\ \hline
 2 &   1   &   9   &   10   &  $-\nicefrac 45$\\ \hline
 3 &   0   &   5   &   5    &  $-1$\\
\end{tabular}
\qquad\qquad\qquad
\begin{tabular}{cV{2.5}c|c|c|c}
$i$ & $r_i$ & $d_i$ & $p_i$ & $\nicefrac{\delta_i}{p_i}$ \\ \hlineB{2.5}
 1 &   8   &   0   &   8    &  1\\ \hline
 2 &   1   &   9   &   10  &  $-\nicefrac 45$\\ \hline
 3 &   0   &   8   &   8    &  $-1$\\
\end{tabular}
\end{center}
For both examples, fix $M=13$. 
We have $K = 2$ in both examples 
because $\delta(D_1) = 8 > 0$ and $\delta(D_2) = 0$.
Both fall under Case 2 because 
 $p(D_1) = 8$ and $p(D_2) = 18$, while $M=13$. 
In the left example there exists an R-majority grouping,
 made by putting together units 1 and 3 to form a grouping
 with $\delta=3$ and 
population 13. But in the right example there is none,
which is easily confirmed by considering all of the combinations.

Finally, in Case 3, we have $p(D_K) < M$.
\begin{claim} Let $S=D_K$ and suppose that $p(S) < M$.
Then for any $S' \subseteq \{1,\dots,n\}$,
$$p(S') > p(S) \implies \delta(S')< \delta(S).$$ 
\end{claim}

The claim asserts that $D_K$ has the optimal R margin among all
groupings with at least as much population.
Since we seek a grouping  larger than $p(D_K)$ and since 
$\delta(D_K) \le 0$, this implies that a R-majority grouping cannot be formed.
So it just remains to prove the claim.

Let $A=S'\setminus S$ and $R=S\setminus S'$ denote the sets of indices
added to and removed from $S$, respectively, to make $S'$.
Since $A$ and $R$ are disjoint, and we have assumed that $p(S')>p(S)$, it follows that $p(A) > p(R)$. 
Let $\mu = \max \{\frac{\delta_i}{p_i} \mid i \in A\}$ and let 
$\mu' = \min \{\frac{\delta_i}{p_i} \mid i \in R \}$. Note that, since $R \subseteq S=\{1,\dots,K\}$ and $A \subseteq 
S^c=\{K\!+\!1,\dots,n\}$ and the $\frac{\delta_i}{p_i}$ are non-increasing, we have 
$\mu\le \mu'$. 

Note that every unit $i \not \in S$ has a Democratic majority $(\delta_i < 0)$. This is because Republican-majority units are added to $S$ in decreasing order of 
$\frac{\delta_i}{p_i}$ until the 
overall $\delta\le 0$, so by construction
every unit with a Republican majority is in $S$.  It follows,
since $A \subseteq S^c$, that $\mu<0$.

We have 
$\mu \cdot p(R) > \mu \cdot p(A)$  because $p(R) < p(A)$ and  $\mu<0$. 
Also,  $\mu' \cdot p(R) \geq \mu \cdot p(R)$. 
So, transitively, $\mu'\cdot p(R) > \mu \cdot p(A) $. 

Note that $$ \mu'\cdot p(R)  = \sum_{i \in R} \mu'\cdot p_i
\leq   \sum_{i \in R} \frac{\delta_i}{p_i} \cdot p_i = \delta(R).$$ 
Similarly $\mu \cdot p(A)  \geq \delta(A)$. 
Combining our inequalities, we have shown that 
 $\delta(R) > \delta(A)$.
It follows that $\delta(S) > \delta(S')$, as claimed.
This completes the proof of the claim and the theorem. 
\end{proof}

Note that Case 2, the inconclusive situation, is more likely when there are units that are large relative to the population threshold,
because the gap between $p(D_{K-1})$ and $p(D_K)$ is the population
of the $K$th unit.
So if we consider the formation of districts,
we are more likely to get an inconclusive result with 
large units like counties or towns and less likely with smaller units like blocks or VTDs/precincts.

This theorem suggests an algorithm for computing feasibility bounds that is no more complex than sorting, which makes it fast and efficient.
The answers are not completely satisfying, however, because of the possibility of an inconclusive finding (Case 2) and because
the existence of a grouping with an R majority and population that is $m$ times the size of an ideal district does not imply that it can be split
into $m$ sub-groupings of equal size, each with R majorities.  However, a refined algorithm that could close those loopholes is known 
to have forbidding computational complexity, because it is equivalent to the {\em $0-1$ knapsack problem}, which is known to 
be NP-complete. \footnote{\url{https://en.wikipedia.org/wiki/Knapsack_problem\#Definition}}

\end{document}